\documentclass[doublecol,figures]{epl2}%
\usepackage{amsmath}
\usepackage{amsfonts}
\usepackage{amssymb}
\usepackage{graphicx}
\usepackage{verbatim}%
\providecommand{\U}[1]{\protect\rule{.1in}{.1in}}

\shorttitle{Crunching Rings} \shortauthor{Julien Fierling}
\institute{ \inst{1} Institut Charles Sadron, CNRS-UdS, 23 Rue du
Loess, BP 84047, 67034 Strasbourg cedex 2, France;\, \\ \inst{2}
Equipe BioPhysStat, LCP-A2MC, Universit\'e de Lorraine, 1 boulevard
Arago, 57070 Metz, France. }  \abstract{ We discuss a
curious example for the collective mechanical behavior of coupled
non-linear monomer units entrapped in a circular filament. Within
a simple model we elucidate how multistability of monomer units
and exponentially large degeneracy of the filament's ground state
emerge as a collective feature of the closed filament.
Surprisingly, increasing the monomer frustration, \textit{i.e.},
the bending prestrain within the circular filament, leads to a
conformational softening of the system. The phenomenon, that we
term polymorphic crunching, is discussed and applied to a possible
scenario for membrane tube deformation by switchable dynamin or
FtsZ filaments. We find an important role of cooperative
inter-unit interaction for efficient ring induced membrane
fission.}
\begin{document}

\title{Crunching Biofilament Rings}
\author{Julien Fierling\inst{1}, Martin Michael M\"{u}ller\inst{1,2}, Herv\'{e}
Mohrbach\inst{1,2}, Albert Johner \inst{1} and Igor M. Kuli\'{c}\inst{1}}
\date{\today }
\maketitle

\section{Introduction}

The classical theme of complex systems centers on the paradigmatic question of
how intricate behavior emerges when elementary units assemble and interact
with each other. In particular, new phenomena can emerge when the elementary
units can switch between several configurational states. Biophysics has
developed a keen interest in these kind of systems as switchable multistable
filaments are found everywhere in living nature. The list of examples is close
to innumerable with the most prominent ones: FtsZ~\cite{FtsZ,FtsZInVitro},
Mrb~\cite{Mrb}, actin~\cite{JanmeyActinRings,SanchezActinRings} or bacterial
flagella~\cite{kamiya,hasegawa}. More recently, microtubules were suggested to
spontaneously form large scale superhelices~\cite{venier, Mohrbach} and
undergo unusual cooperative dynamics \cite{Osman}. Even whole microorganisms
exhibit switchability inherited from their constituent
filaments~\cite{HelicalBacteria,Arroyo}.

In this Letter we study an interesting example for the emergence of an
unexpected behavior when non-linear units are entrapped in a circular
(bio)filament. More precisely, we pose the following conceptual question: What
happens when protein monomers with an intrinsic curvature form a stiff polymer
which is forced to close in a ring of different curvature radius? Biofilament
rings and shallow helices are a common theme in biological actuation and have
been in the focus of numerous experimental studies
\cite{FtsZ,FtsZInVitro,JanmeyActinRings,SanchezActinRings}.

As we will see in this Letter, the closure constraint can turn the monomers
into fluctuating bistable units with rather dramatic effects on the overall
shape of the ring. We find that the ground state is not unique but extremely
degenerate. Its number of realizations increases exponentially with the number
of monomers $N$. As a consequence, increasing the length of the chain $L$
leads to a conformational softening of the closed filament. We call this
phenomenon of extreme shape degeneracy \textquotedblleft polymorphic
crunching\textquotedblright. In the second part of this Letter we investigate
the effects of monomer cooperativity and study how a crunching polymorphic
ring could constrict a membrane tube and lead to membrane fission. We show
that a filament ring can spontaneously crunch even in the absence of
cooperativity. However, it can only constrict the membrane in the presence of
strong enough inter-monomer cooperativity.

\section{Emergence of multistability}

\begin{figure}[t]
\centering
\includegraphics* [ width=8.5cm]{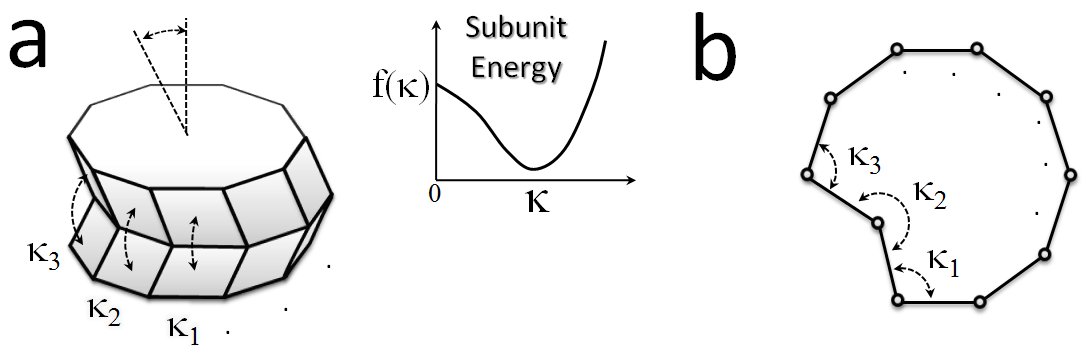} \caption{Two basic types of
collective coupling between non-linear units, with curvature $\kappa_{i}$ and
bending energy $f(\kappa_{i})$, via a global constraint. a) Polymorphic
buckling as found in bacterial flagella and microtubules. Bendable units are
coupled in the plane \textit{perpendicular} to their bending axis
\cite{Mohrbach,Osman}. b) Polymorphic crunching, considered in this Letter.
Nonlinear bendable units are coupled \textit{in-plane of bending} by a ring
closure constraint.}%
\label{fig:image1}%
\end{figure}

To grasp the flavor of the problem, let us investigate how switchability and
multistability of coupled units arise in two simple but generic situations,
cf. Fig. 1. For this, consider a collection of identical elastic elements -
monomers - each one of them characterized by its curvature $\kappa_{i}.$ We
assume further that each monomer has a bending energy $f\left(  \kappa
_{i}\right)  $ where the function $f$ has a global minimum at $\kappa
=\kappa_{m}$ corresponding to the \textit{monomer's intrinsic curvature, }cf.
Fig 1a, right panel.

As a first example let us consider a situation where $N$ such monomer units
are mechanically coupled in the plane \textit{perpendicular} to their bending
axis (see Fig.~\ref{fig:image1}a). In this manner the monomers form a short
slice of a tube of circular cross-section. When the tube section bends in a
direction orthogonal to the cross-section and assumes a curvature
$\kappa_{tube}$, the monomers themselves adapt accordingly and become curved
with $\kappa_{i}\approx\kappa_{tube}\sin\frac{2\pi i}{N}$ as a result of the
geometric projection of the tube's curvature on each monomer unit. The total
elastic energy of a tube in this geometry is then given by \cite{Mohrbach}%
\begin{equation}
E_{tube}\left(  \kappa_{tube}\right)  =\sum_{i=1}^{N}f\left(  \kappa
_{tube}\sin\frac{2\pi i}{N}\right)  \;.
\end{equation}
In the simplest case of $N=2$, \textit{i.e.}, only two monomers
\textquotedblleft welded\textquotedblright\ together the total energy reads
$E_{tube}(\kappa_{tube})=f(\kappa_{tube})+f(-\kappa_{tube}),$ where for
simplicity we assume curvature perpendicular to the welding line. Due to the
mirror symmetry of the problem, the total energy depends only on even terms of
$f$. In the most general case, we expect the system to admit at least three
equilibrium configurations: two symmetric bent states looking like
$\langle\langle$ and $\rangle\rangle$ as well as a straight state $||$. The
straight state $||$ becomes unstable when $\left.  \frac{\partial^{2}%
f}{\partial^{2}\kappa}\right\vert _{\kappa=0}<0,$ leading to two (in general
$N$) equivalent ground states $\langle\langle$ and $\rangle\rangle$. This
symmetry breaking and the emergent bi/multi- stability is the basic motif in
switchable tubular systems (see Fig.~\ref{fig:image1}a) such as microtubules
\cite{Mohrbach,Osman} and bacterial flagella \cite{Asakura, Calladine, Powers,
Netz}.

In a second example, let us consider an even simpler coupling of monomer
units:\ \textit{in the plane} of their bending degree of freedom (see
Fig.~\ref{fig:image1}b). What happens when a chain of such monomers is forced
to close into a planar ring? As in the discussion above let us start with a
model-independent general analysis of such a ring. In two dimensions the
conformation of such a filament of length $L$ consisting of $N=L/a$ identical
monomers of size $a$ is given by its signed curvature $\frac{d\theta}%
{ds}=\kappa\left(  s\right)  $ where $s\in\lbrack0,L]$ is the arc-length
variable and $\theta$ the tangent angle of the filament's centerline. Again,
as in the previous example, each unit has an elastic energy $f\left(
\kappa\right)  .$ In the present case the total ring energy reads%
\begin{equation}
E_{ring}=\int_{0}^{L}f\left[  \kappa\left(  s\right)  \right]  ds\;.
\label{EringTotal}%
\end{equation}
For a linearly elastic polymer ring with elastic energy $f\left(
\kappa\right)  \propto\kappa^{2}$ the ground state would be a perfect circle
of curvature $\kappa_{0}=2\pi/L$. What happens for strongly nonlinear
$f\left(  \kappa\right)  ?$ For small deviations $\delta\kappa$ from the
circle the curvature can be decomposed: $\kappa\left(  s\right)  =\kappa
_{0}+\delta\kappa\left(  s\right)  $ and we can expand the elastic energy
\begin{equation}
E_{ring}=E_{0}+\sum_{n\geq2}\left.  \frac{1}{n!}\frac{\partial^{n}f}%
{\partial\kappa^{n}}\right\vert _{\kappa=\kappa_{0}}\,\int_{0}^{L}%
\!\!\delta\kappa^{n}\text{d}s\;,
\end{equation}
where $E_{0}$ is the energy of the perfect circle. The sum starts at $n=2$ due
to the tangent constraint $\int_{0}^{L}\!\kappa ds=2\pi$ implying $\int
_{0}^{L}\!\delta\kappa\,ds=0$.

We see that the circle is locally stable if $\frac{\partial^{2}f}%
{\partial\kappa^{2}}|_{\kappa=\kappa_{0}}>0$. In the opposite case,
$\frac{\partial^{2}f}{\partial\kappa^{2}}|_{\kappa=\kappa_{0}}<0$, the
circular ring becomes unstable. One of the main goals of this letter is to
study the ground state for such a polymorphic ring.
For the sake of concreteness let us assume that each monomer has a bending
energy per length given by:%
\begin{equation}
f(\kappa)=A\kappa-\frac{B}{2}\kappa^{2}+\frac{C}{2}\kappa^{4}\;, \label{Ering}%
\end{equation}
where $A$, $B>0$, and $C>0$ are elastic constants whose signs are chosen in
such a way that at least one stable preferred curvature state exists. The
constant $A$ represents a structural \textit{asymmetry} of the monomer
unit\cite{NoteCubicTerm}. An open filament described by Eq.~(\ref{Ering}),
behaves like an intrinsically curved ribbon with two distinguishable inner and
outer faces. For $A$ large enough the monomer has only one preferred state
given by $\kappa\approx-\left(  \frac{A}{2C}\right)  ^{1/3}$.

Remarkably, if we now \textit{close} the filament, the asymmetric term
$A\int_{0}^{L}\kappa\left(  s\right)  ds=A(\theta\left(  L\right)
-\theta\left(  0\right)  )=2\pi A$ \ gives merely a constant in the ring
energy Eq.~(\ref{EringTotal}) and can thus be discarded. We end up with an
\textit{effective} total energy of the ring:
\begin{align}
E  &  =\int_{0}^{L}f_{eff}\left(  \kappa\right)  ds,\qquad\text{where}%
\label{Eeff}\\
f_{eff}\left(  \kappa\right)   &  =-\frac{B}{2}\kappa^{2}+\frac{C}{2}%
\kappa^{4}\;. \label{Eff2}%
\end{align}
The closure condition implies an invariance with respect to the choice of the
monomer asymmetry $A$ and has interesting physical implications. In the case
where the \textit{free} monomer has only a single preferred state the
constraint induced by chain closure makes the monomer effectively bistable.
Indeed, the corresponding effective potential $f_{eff}(\kappa)$ (an even
function) has always two equivalent minima given by $\kappa=\pm\kappa_{1}$
with the characteristic curvature
\begin{equation}
\kappa_{1}=\sqrt{\frac{B}{2C}}%
\end{equation}
and $f_{eff}\ (\kappa_{1})=-\frac{1}{8}\frac{B^{2}}{C}.$ Remarkably it is the
chain closure, \textit{i.e.}, a global topological property of the system that
generates an \textit{emergent bistability} of the local constituent.

If each monomer is close to its ground state , $\kappa(s)=\pm\kappa_{1}%
+\delta\kappa_{el}(s),$ with some additional elastic thermal fluctuations
$\delta\kappa_{el}(s)$, we obtain the total elastic energy up to quadratic
order:
\begin{equation}
E_{elastic}\ =-\frac{L}{8}\frac{B^{2}}{C}+B\int_{0}^{L}\delta\kappa_{el}%
^{2}ds\;. \label{Eelas}%
\end{equation}
From this result one could na\"ively infer the associated persistence length
of the system to be $l_{B}=2B/(k_{B}T)$. However, a second contribution of a
purely entropic origin due to the degeneracy of the ground state has to be
taken into account as we will show in the following.

\section{The crunching transition}

\begin{figure}[t]
\centering
\includegraphics* [width=8.5cm]{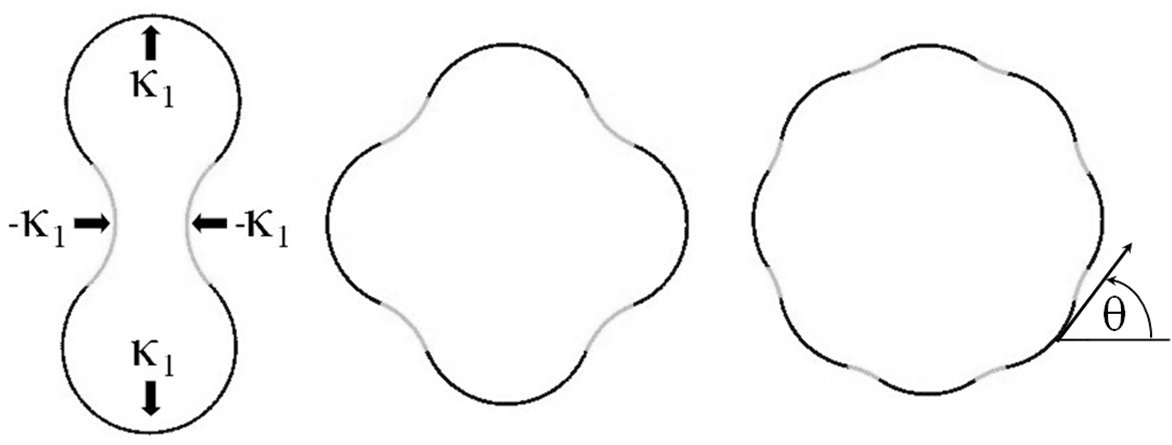} \caption{ Crunching of a circularly
closed non-linearly elastic filament. A ground state consists of a sequence of
positive and negative curvature regions. All depicted states are ground states
and have the same elastic energy.}%
\label{fig:image2}%
\end{figure}

The emergence of the two stable monomer states leads to the
conclusion that the ground state of the system defined by
Eq.~(\ref{Eeff}) is highly degenerate. At vanishing temperature,
and once the parameter $\kappa _{1}/\kappa_{0}$ surpasses $1,$ any
curvature function of the form $\kappa\left(  s\right)
=\pm\kappa_{1}$ gives the same minimum ring energy. In the
discrete notation the curvature of the $n$th monomer can be
written as $\kappa_{n}=\sigma_{n}\kappa_{1}$ with the "spins"
$\sigma_{n}=\pm1$, $n=1,\ldots N$. For $\kappa_{1}/\kappa_{0}>1$
exponentially many equivalent ground-states with different shapes
become possible. The onset of this sudden behavior at $T=0$, we in
the following call the crunching transition. Does the sharp onset
of the crunching transition persist or does it get "washed-out" by
the fluctuations at $T>0$? Let us investigate this question in the
following.

For a hypothetically open filament described by Eq.~(\ref{Eeff})---in the
absence of the closure condition---the system is formally equivalent to a
one-dimensional Ising chain without interactions between the spins. In analogy
to the latter, at finite temperature $T$ we introduce the "average
magnetization"
\begin{equation}
M=\left\langle \frac{1}{N}\sum_{n=1}^{N}\sigma_{n}\right\rangle \;,
\end{equation}
describing the mean of the distribution of positively and negatively curved
monomers with the average curvature $\left\langle \kappa_{n}\right\rangle
=\kappa_{1}M$ . For a given value of $M$, the free energy $F(M)=E-TS(M)$
decomposes into the elastic energy $E=-\frac{L}{8}\frac{B^{2}}{C}$ and the
large entropic contribution $S(M)=k_{B}\ln W(M)$ due to the high degeneracy of
the ground state\cite{NoteElastic}. Here $W(M)$ is the number of ring ground
states for a given $M$. The physical value of $M$ at any temperature is given
by the minimum of $F(M)$ which is obviously $M=0$ for the open filament
defined by Eq.~(\ref{Eeff}). This implies that the ground states, although
degenerate and locally tortuously bent (between -$\kappa_{1}$ and +$\kappa
_{1}$ regions) are however statistically straight on average. Note that this
2-state model bears some resemblance with other bistable systems like the
squeezed helices studied in Ref~\cite{Nam}. It can also can be seen as a
concrete realization of the block-copolymer model for curvature switchable
filaments developed in Ref.~\cite{JulienEPL}.

Closing the filament imposes three separate constraints. First and second the
closure constraint in the ($xy$)-plane:
\begin{equation}
\int_{0}^{L}\cos\theta\left(  s\right)  ds=\int_{0}^{L}\sin\theta\left(
s\right)  ds=0 \label{X-Y-Closure}%
\end{equation}
and third, the condition $\theta\left(  L\right)  =2\pi+\theta\left(
0\right)  $, resulting from the continuity of the tangent angle. In order to
adapt to these constraints, the curvature distribution of the monomers along
the filament will not be random any more. Indeed the non-linear closure
constraints Eq.~(\ref{X-Y-Closure}) introduce a weak non-local effective
coupling between the individual monomers. Besides, there is now a bias of mean
curvature in favor of the curvature $\kappa_{0}$, i.e. $\left\langle
\kappa_{n}\right\rangle =\kappa_{0}$, necessary for ring closure.

Nevertheless, similar to the previous case of the open filament the ground
state shows a high level of degeneracy (see Fig.~\ref{fig:image2}). Solving
the problem with the non-linear closure constraints Eq.~(\ref{X-Y-Closure}) is
a rather difficult task, but to illustrate the idea of the ground state
degeneracy we consider a simple case: If the curvature $\kappa\left(
s\right)  $ displays a four-fold symmetry with respect to the $x$ and $y$ axes
then the strict geometric closure is ensured by symmetry. In this case we can
restrict ourselves to one of the four quarter copies of the filament with
$N/4$ monomers each.

In contrast to the open chain, once the ring is geometrically closed, it seems
that crunching could happen only for $\kappa_{1}>\kappa_{0}.$ For $\kappa
_{1}\leq\kappa_{0}$ only the perfect circle seems admissible. However the
monomers need not necessarily keep the curvature $\pm\kappa_{1}$ but can move
slightly away from the minimum of the effective potential~(\ref{Eff2}). Indeed
such a shift of the curvature, although energetically costly can be
compensated by the entropic contribution. A small additional curvature
$\Delta_{\kappa}\kappa_{1}$ (away from the mechanical ground state) added to
each monomer $\kappa_{n}=\left(  \sigma_{n}+\Delta_{\kappa}\right)  \kappa
_{1}$ helps the system to reach macroscopic states with more configurational
realizations then in the strict ground state. With this variational ansatz for
the constrained ground state(s) the angular closure constraint implies
\begin{equation}
M=\frac{4}{N}\left\langle \sum_{n=1}^{N/4}\sigma_{n}\right\rangle
=\frac{\kappa_{0}}{\kappa_{1}}-\Delta_{\kappa}\; , \label{KappaConstr}%
\end{equation}
where $0<M<1$. Note that in the case where $\kappa_{1}<\kappa_{0}$,
$\Delta_{\kappa}$ is necessarily non zero in order for $M<1$. With this ansatz
the statistical average of the curvature is by construction $\left\langle
\kappa_{n}\right\rangle =\kappa_{0}$. Again the free energy $F=E-TS$ of the
system contains the elastic energy $E(\Delta_{\kappa})$ and the entropic
contribution $S(\Delta_{\kappa})=k_{B}\ln W(\Delta_{\kappa})$ due the high
degeneracy of the ground state. Now $W(\Delta_{\kappa})$ is the number of
states compatible with the condition~(\ref{KappaConstr}). For a finite $N$ and
fixed value of the other parameters $\Delta_{\kappa}$ varies with temperature
$T$ and $N$. When the energy is dominating over the entropy (small $T$ or
small $N$) $\Delta_{\kappa}\approx0$.

Let us consider the case of $N$ large with $L=Na$ fixed and $a$ the size of
the monomer. With the notation $\tilde{\kappa}_{1}=\kappa_{1}/\kappa_{0}$ the
energy density can be written as:
\begin{equation}
\frac{E}{aBN\kappa_{0}^{2}\tilde{\kappa}_{1}^{2}}=-\frac{1}{4}+\Delta_{\kappa
}^{2}\left(  1+\frac{\Delta_{\kappa}}{\tilde{\kappa}_{1}}-\frac{3}{4}%
\Delta_{\kappa}^{2}\right)
\end{equation}
and the entropy per monomer becomes
\begin{align}
\frac{S}{k_{B}N}  &  =\ln2-\frac{1}{2}(1-\Delta_{\kappa}+\frac{1}%
{\tilde{\kappa}_{1}})\ln(1-\Delta_{\kappa}+\frac{1}{\tilde{\kappa}_{1}%
})\nonumber\\
&  -\frac{1}{2}(1+\Delta_{\kappa}-\frac{1}{\tilde{\kappa}_{1}})\ln
(1+\Delta_{\kappa}-\frac{1}{\tilde{\kappa}_{1}}).
\end{align}
The minimization of $F$ with respect to $\Delta_{\kappa}$ leads to the
relation:
\begin{equation}
\Delta_{\kappa}=\frac{1-\tilde{\kappa}_{1}+(1+\tilde{\kappa}_{1}%
)\exp(-\varepsilon)}{\tilde{\kappa}_{1}\left(  1+\exp(-\varepsilon)\right)
}\;,
\end{equation}
where $\varepsilon$ is given by
\begin{equation}
\varepsilon=4al_{B}\kappa_{0}^{2}\tilde{\kappa}_{1}^{2}\left(  2-3\Delta
_{\kappa}^{2}+3\frac{\Delta_{\kappa}}{\tilde{\kappa}_{1}}\right)
\Delta_{\kappa}%
\end{equation}

For a very stiff filament whose bending stiffness $B=l_{B}kT/2\gg kT/\left(
a\kappa_{0}^{2}\right)  $ dominates over the entropy, the value of
$\Delta_{\kappa}$ jumps sharply from $\Delta_{\kappa}=1/\tilde{\kappa}_{1}%
$\ (for $\tilde{\kappa}_{1}<1$) to $\Delta_{\kappa}=0$ (for $\tilde{\kappa
}_{1}>1$) close to the "crunching transition point" $\tilde{\kappa}_{1}=1$.
Consequently, the filament stays a ring with $\kappa=\kappa_{0}$ below the
transition $\left(  \tilde{\kappa}_{1}<1\right)  $ and is crunched above it.
For more moderate stiffness $B\lesssim kT/\left(  a\kappa_{0}^{2}\right)  $,
the system more gradually interpolates between the uncrunched and crunched
state and the transition smoothens increasingly with decreasing $B$.

The lesson from this mean-field analysis is that in general the local
curvature $\kappa_{n}$ can deviate from its preferred value ($\pm\kappa_{1}$)
by an amount $\Delta_{\kappa}\kappa_{1}$. Classically, at $T=0$, the crunching
happens only once $\kappa_{1}$ surpasses $\kappa_{0}.$ However, in the
presence of an additional elastic deformation $\Delta_{\kappa}$ the filament
can "pre-crunch" even for $\kappa_{1}<\kappa_{0}$. The exponentially large
number of curved state micro-realizations (reflected in the entropy $S\left(
\Delta_{\kappa}\right)  \propto N\ $scaling with the system size)\ competes
with the elastic energy $E\left(  \Delta_{\kappa}\right)  $ caused by the
additional deformation $\Delta_{\kappa}.$ This effectively smoothens and
prevents a strict thermodynamic phase-transition at $\kappa_{1}=\kappa_{0}$
(occuring at $T=0$) at any finite $T>0.$

\section{Softening Through Prestrain}

How does the presence of exponentially many equivalent states
affect the shape fluctuations of a closed ring? From the previous
analysis we can now make a coarse-grained approach in the limit of
large number of bistable monomers, $N\gg1$. We further assume that
the curvature $\kappa_{1}>\kappa_{0}$ is supercritical , which
gives rise to a negligible $\Delta_{\kappa}$ from previous
section. As already mentioned, the closure
conditions~(\ref{X-Y-Closure}) result in a weak non-local coupling
between the individual monomers. This closure coupling, which we
previously circumvented by assuming 4-fold symmetry, will be
treated more elegantly later on by eliminating certain Fourier
modes in the coarse-grained filament description. For now, we
neglect the strict closure condition and impose the ring
conformation on average: $\left\langle \kappa_{n}\right\rangle
=\kappa_{0}$. In this way the spins $\sigma_{n}$ can be considered
as randomly distributed around an average value $\left\langle
\sigma_{n}\right\rangle =M$ (Eq.~(\ref{KappaConstr})).

\begin{figure}[t]
\centering
\includegraphics* [width=6.5cm]{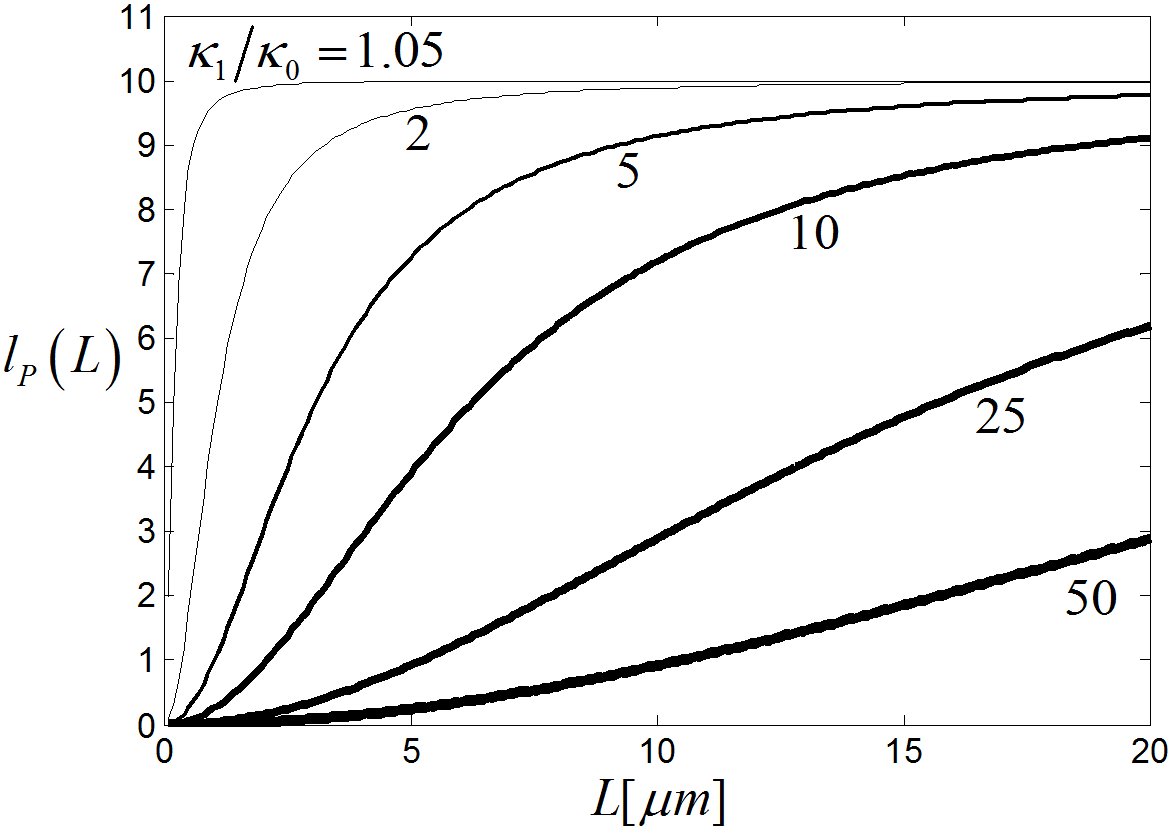} \caption{The persistence length
$l_{p}$ of crunched filament rings depends on their length
(Eq.~(\ref{eq:totalperslength})). Parameters, $a=1\text{nm}$, $l_{B}%
=10\mu\text{m}$, $\kappa_{0}=2\pi/L\ $and an increasing polymorphic curvature
$\kappa_{1}$.}%
\label{fig:image3}%
\end{figure}

For a large number of monomers the piecewise function $\kappa_{n}=\pm
\kappa_{1}$ satisfies the central limit theorem so that the probability to
find a certain curvature $\left\langle \kappa_{n}\right\rangle _{l}$
appropriately coarse-grained on a larger lengthscale $l\ll Na$ (between the
monomer size and the total ring size) is a gaussian variable. Using this
coarse-grained approach, valid on scales $l\gg a$ beyond the monomer size, we
treat the curvature as a continuous gaussian-distributed function $\kappa(s)$.
We now introduce a new random gaussian variable $\tilde{\kappa}(s)=\kappa
(s)-\left\langle \kappa(s)\right\rangle =\kappa(s)-\kappa_{0}$ with zero mean
and a standard deviation given by $\left\langle \tilde{\kappa}(s)^{2}%
\right\rangle =\kappa_{1}^{2}-\kappa_{0}^{2}\overset{(\ref{KappaConstr})}%
{=}\kappa_{1}^{2}(1-M^{2})$. Since the curvature is uncorrelated on larger
scales, $\left\langle \tilde{\kappa}(s)\tilde{\kappa}(s^{\prime})\right\rangle
\approx\delta\left(  s-s^{\prime}\right)  \left\langle \tilde{\kappa}%
(s)^{2}\right\rangle $ for $\left\vert s-s^{\prime}\right\vert \gg l$. Recast
in this form, the polymorphic chain can now be seen in a new light: The
coarse-grained curvature $\tilde{\kappa}(s)$ behaves on larger scales as the
curvature of a classical semiflexible chain. The latter is described by the
Worm-Like Chain (WLC) model with an effective configurational energy:
\begin{equation}
E_{WLC}=\frac{l_{e}k_{B}T}{2}\int_{0}^{L}\tilde{\kappa}(s)^{2}\text{d}s
\label{WLC}%
\end{equation}
and an effective persistence length given by
\begin{equation}
l_{e}=\frac{1}{a(\kappa_{1}^{2}-\kappa_{0}^{2})}\;. \label{eq:effperslength}%
\end{equation}
Despite this useful analogy with the WLC a note of caution is appropriate.
Unlike for the WLC, the persistence length $l_{e}$ depends here on the length
$L$ of the filament through $\kappa_{0}=2\pi/L$ and is moreover
temperature-independent since it is associated to the conformational entropy
of the ground state. From Eq.~(\ref{eq:effperslength}) we see that $l_{e}$ is
a decreasing function of $L$. The chain softens for increasing length until
$l_{e}$ reaches its minimum value $l_{e}=1/\left(  a\kappa_{1}^{2}\right)  $.

Until now we implemented only the mean condition $\left\langle \kappa
_{n}\right\rangle =\kappa_{0}$. To enforce the closure
conditions~(\ref{X-Y-Closure}), consider the Fourier decomposition of the
tangent angle $\theta$ around a circular state, \textit{i.e.}, $\theta
(s)=\kappa_{0}\,s+\delta\theta(s)$, where $\delta\theta(s)$ is small:
$\delta\theta(s)=\sum_{n=1}^{N}\left(  a_{n}^{\theta}\sin(\kappa_{0}%
ns)+b_{n}^{\theta}\cos(\kappa_{0}ns)\right)  $. Expanding the X-Y
constraints~(\ref{X-Y-Closure}) to lowest order in modes $a_{n}^{\theta}$ and
$b_{n}^{\theta},$ one immediately sees that $a_{1}^{\theta}$ and
$b_{1}^{\theta}$ both have to vanish. The higher modes are unaffected and
according to the equipartition theorem satisfy: $\left\langle (a_{n}^{\theta
})^{2}\right\rangle =\left\langle (b_{n}^{\theta})^{2}\right\rangle =\left(
L/2\pi^{2}l_{e}\right)  n^{-2}\;,n\geq2$ which are temperature-independent due
to purely entropic effects.

Including the elastic thermal fluctuations around the ground state(s), it is
easy to see that the total persistence length reads
\begin{equation}
l_{p}\left(  L\right)  =\frac{l_{B}l_{e}(L)}{l_{B}+l_{e}(L)}\;
\label{eq:totalperslength}%
\end{equation}
which is always smaller than the persistence length $l_{B}$ (see
Fig.~\ref{fig:image3}). We thus observe an effective conformational softening
of the filament due to the prestrain induced by the closure. In particular,
when the temperature goes to zero or when $l_{B}$ is very large (such that
$l_{B}\gg l_{e}$), $l_{p}$ does not diverge as it would for a standard
semiflexible filament but stays finite, i.e. $l_{p}=l_{e}(L)$.

\section{Cooperativity}

Many biofilaments, including dynamin, FtsZ, tubulin and others are known to
switch cooperatively. To incorporate this effect, in addition to the elastic
energy Eq.~(\ref{Eeff}) we assume a cooperative inter-monomer coupling term
favoring uniform curvature%
\begin{equation}
E_{couple}=\frac{K}{2}\int_{0}^{L}\left(  \frac{d\kappa}{ds}\right)  ^{2}ds
\label{Ecouple}%
\end{equation}
with an \textit{inter-monomer coupling} constant $K.$ With cooperativity,
domains switch between $\pm\kappa_{1}$ on a scale $\lambda\sim\sqrt{K/B}$ and
have transition energy penalty $\ J\sim\sqrt{KB}\kappa_{1}^{2}.$ The effective
block size $\xi\sim ae^{J/kT}$ of $\pm\kappa_{1}$ blocks can now be considered
as the effective monomer size $a_{eff}\sim\xi.$ For short transition lengths
$\lambda\ll\xi$ and intermediate cooperativity, $a<a_{eff}\sim\xi\ll L$, all
the results from the previous sections stay applicable, however with a
\textit{renormalized monomer size} $a_{eff}$. In particular the system behaves
again like a WLC on larger scales but with smaller persistence length
$l_{e}\propto1/a_{eff}$.\ For very large $\xi\gg L$ (in practice
$J\gtrsim5-10kT$) the domain walls become too costly and $a_{eff}>L$ , with
only a minimal number of walls being energetically admissible. In this highly
cooperative limit the ring cannot be reduced to a fluctuating WLC, but has a
single well defined ground state, assuming a definite shape with 4-domains as
Fig. 2, left panel. This shape will be preferred over the simple circle if the
energy of the four transition regions becomes less than the penalty for having
a uniform curvature $\kappa_{0}$ i.e. $4J\lesssim\frac{LB^{2}}{8C}%
=BL\kappa_{1}^{2}/4.$

\section{Filament Crunching a Membrane Tube}

\begin{figure}[t]
\centering
\includegraphics* [width=7.5cm]{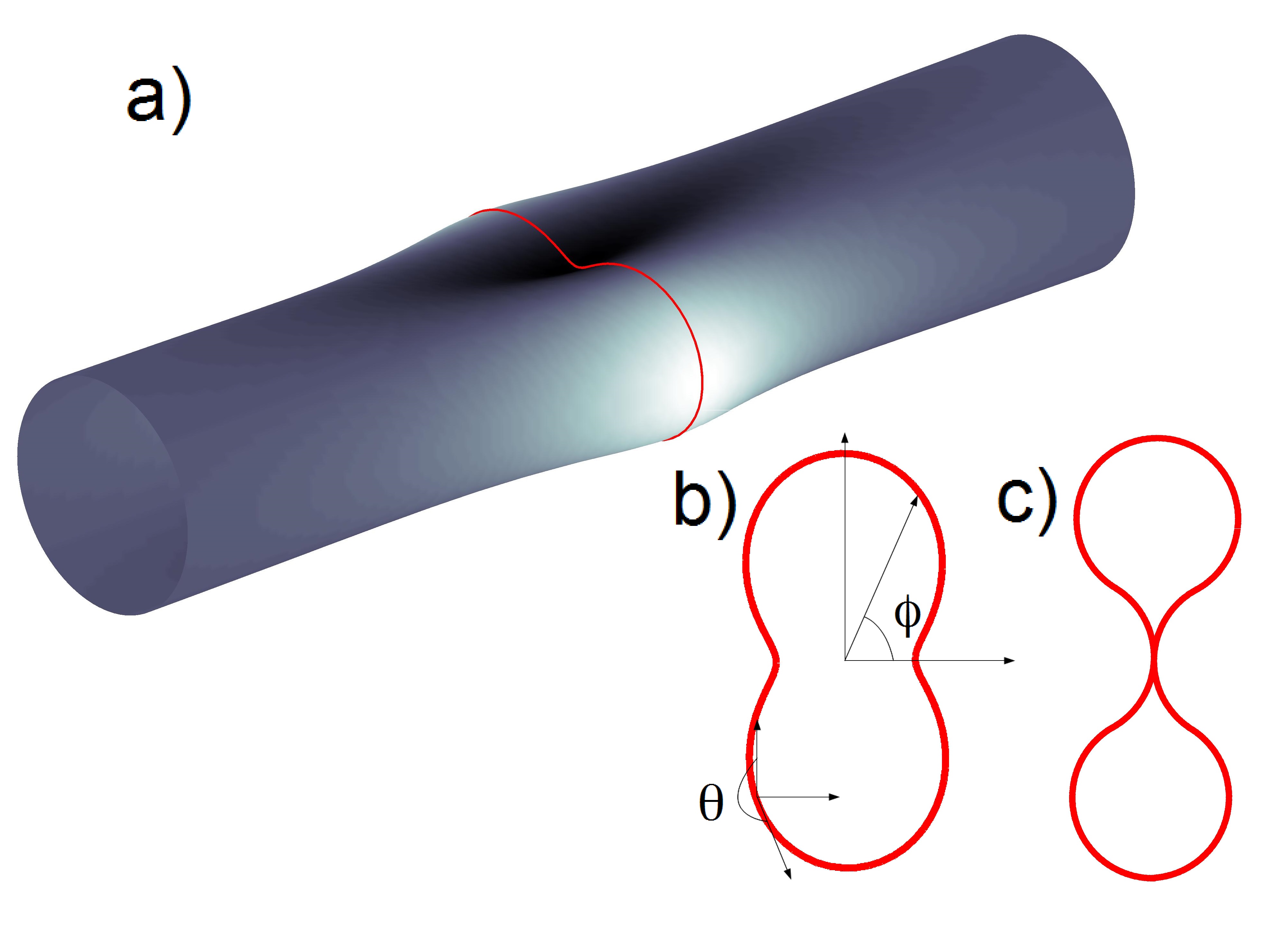}\caption{a) A crunching polymer ring
on a fluid membrane, b) weakly deformed, c) critical deformation with self
contact for $\kappa_{1}/\kappa_{0}=7/3$. }%
\label{fig:image4}%
\end{figure}

In order to illustrate the model, we apply it to a possible scenario for
membrane tube crushing by a polymorphic filament. In nature one finds many
examples of biological polymers interacting with tubular membranes. For
instance, the dynamin filament wraps around a neuron's membrane during
endocytosis \cite{DynaminRoux,DynaminJoanny} to form a helix whose pitch is
small. FtsZ filament rings constrict the membranes of tubular bacteria during
cell division \cite{FtsZ,FtsZInVitro}. We do not pretend to specifically model
dynamin or FtsZ in this paper but rather set the framework to better
understand such interactions in the near future.

We consider a tubular fluid membrane of infinite length, radius $r$ and
bending rigidity $B_{t}$. In terms of the azimuthal angle $\phi$, the
longitudinal coordinate $z$, and the radial displacement field $u(\phi,z)$
with respect to the cylindrical state the elastic energy reads
\cite{Parmeggiani} $F_{t}=\int\int_{0}^{2\pi}e(\phi,z)$d$\phi$d$z$ with
$e(\phi,z)=B_{t}[(\partial_{\phi}^{2}u)^{2}+3u^{2}+2r^{2}(\partial_{z}%
^{2}u)(\partial_{\phi}^{2}u)-2r^{2}(\partial_{z}u)^{2}+r^{4}(\partial_{z}%
^{2}u)^{2}+4u(\partial_{\phi}^{2}u)]/2r^{3}$. Fourier decomposition of
$u(\phi,z)$ reads:
\begin{equation}
u(\phi,z)=\sum\nolimits_{n}^{^{\prime}}\int\frac{dk}{2\pi}A_{n}(k)\text{e}%
^{i(n\phi+kz)}\;, \label{eq:decompositionmembrane}%
\end{equation}
where in the sum $\sum^{\prime}$ the modes $n=\pm1$ (for all $k$) are
excluded. They would correspond to a net external force, which is absent in
our problem. The associated quadratic energy reads \cite{Parmeggiani}:
\begin{equation}
F_{t}=\frac{\pi B_{t}}{r^{3}}\sum\nolimits_{n}^{^{\prime}}\int\frac{dk}{2\pi
}\mathcal{M}_{n}(k)A_{n}(k)A_{n}(k)^{\ast}\; \label{eq:energymembrane}%
\end{equation}
with $\mathcal{M}_{n}(k)=(r^{2}k^{2}+n^{2}-1)^{2}-2(n^{2}-1)$.

When the polymorphic filament wraps around the tube it induces a deformation
at position $z=0$ of the form: $u(\phi,0)=\sum_{n=-\infty}^{\infty}%
a_{n}\text{e}^{in\phi}$. Expressing the membrane modes $A_{n}(k)$ in terms of
the filament modes $a_{n}$ we obtain the energy of the constrained membrane:
\begin{equation}
F_{t}=\frac{\pi B_{t}}{r^{2}}\sum\nolimits_{n}^{^{\prime}}\frac{|a_{n}|^{2}%
}{I_{n}}\;, \label{eq:energymembrane2}%
\end{equation}
where $I_{n}=r\int\frac{dk}{2\pi}\frac{1}{\mathcal{M}_{n}(k)}$. The shape of
the membrane at position can also be easily deduced from filament
modes\cite{NoteMembraneShape}.

Can a polymorphic filament deform or even crunch the membrane tube? Consider
first the case where cooperativity is negligible and the filament behaves like
a WLC on large scales. In the limit of large $l_{B}\gg l_{e}$ we can neglect
the elastic fluctuations and $l_{p}\approx l_{e}$. The total free energy in
this case becomes $F=F_{WLC}+F_{t}$, where $F_{WLC}$ and $F_{t}$ are given by
Eqns.~(\ref{WLC}) and (\ref{eq:energymembrane2}), respectively. In terms of
the Fourier modes $a_{n}$ of the radial displacement $u$, the ring
contribution is $F_{WLC}=\frac{k_{B}Tl_{e}\kappa_{0}^{2}L}{r^{2}}\sum_{n}%
n^{4}|a_{n}|^{2}$. Applying the equipartition theorem to each mode we obtain
$\left\langle |a_{n}|^{2}\right\rangle =2r^{2}\,/\left(  \frac{\pi l_{e}}%
{r}n^{4}+\frac{\pi B_{t}}{I_{n}k_{B}T}\right)  $ for $|n|\neq1$. We note that
the filament always \textit{reduces} the free membrane fluctuations (given by
$l_{e}/r\rightarrow0$). Therefore, a filament which is lacking structural
cooperativity is unable to deform the membrane.

However, the presence of high cooperativity (Eq.\ref{Ecouple}) changes this
picture completely and boosts dramatically the deformation of the membrane. A
very stiff filament $B\gg rB_{t}$ with high cooperativity $J\gg kT$ dominates
entirely over the membrane. For $\kappa_{1}>4\sqrt{J/BL}$ the filament
establishes a peanut-shape, cf. Fig 4. The tube deforms strongly and
establishes a self contact for $\kappa_{1}/\kappa_{0}=7/3$ ("kissing
condition", cf. Fig.~\ref{fig:image4}). Beyond this critical value the
membrane would be forced to tear and rupture due to strong self interaction
and the localized shear induced by the filament slicing through it.

\section{Conclusions}

We have shown that the conceptually simple procedure of circular closure
transforms a simple mundane object---an anharmonic filament with a unique
ground state---into a complex multistable filament with an exponentially large
number of degenerate ground states. When the filament thermically explores
this multistable energy landscape, it exhibits anomalous fluctuations. We have
shown that in the limit of low cooperativity the filament ring can be modelled
as a worm like chain but with an effective (length dependent) persistence
length which is dominated by configurational fluctuations between the ground states.

Motivated by FtsZ and dynamin filaments we have started to explore the
interaction of such a multistable crunching filament with a tubular fluid
membrane. We have seen that a crunching ring can deform a membrane tube only
in the presence of strong inter-monomer cooperativity. In this cooperative
limit the membrane could undergo fission in a novel geometric scenario: The
membrane is forced through itself and possibly ruptured by the crunching
filament slicing through it. The self-contact of the membrane driven by
localized normal pressure along a line (the filament) and the subsequent
recombination pathway of the upper and lower membrane leaflets is an
interesting problem, worthwhile exploring in future.


\begin{thebibliography}{99}
\bibitem {NoteCubicTerm} Note that more generally a term $\propto\kappa^{3}$ is possible
and its presence modifies the emergence of two exactly symmetric
minima. In our later exposition, the assymetry can be incorporated
by a biasing "magnetic" field. For simplicity we focus here on the
simpler case where the cubic and higher odd terms are absent or
negligible.
\bibitem {NoteElastic}The elastic thermal fluctuation contributions
$\delta\kappa_{el}$ around the ground states average to $\left\langle
\delta\kappa_{el}\right\rangle =0$. They only add a constant (entropic) term
to the full free energy of the system and can be disregarded from now.
\bibitem {NoteMembraneShape}
The membrane shape is then given by
$u(\phi,z)=\sum\nolimits_{n}^{^{\prime}}a_{n}$e$^{in\phi}J_{n}(z)/I_{n}$
with $J_{n}(z)=\left(  8\sqrt{2}\sqrt {n^{2}-1}\right)
^{-1}\left(
e^{-\frac{z}{r}\gamma^{-}}/\gamma^{-}-e^{-\frac{z}{r}\gamma^{+}}/\gamma^{+}\right)
$ and $\gamma^{\pm}=\left( \sqrt{n^{2}-1}\left(
\sqrt{n^{2}-1}\pm\sqrt{2}\right)  \right)  ^{1/2}$,
$I_{n}=J_{n}(0)$ for $|n|\neq1$ and $I_{0}\approx0.06$.
\end{thebibliography}
\end{document}